\documentclass[prb,twocolumn,aps,showpacs]{revtex4}
\usepackage{graphicx}
\def\ltsim{\vbox {\hbox{\lower .8\baselineskip \hbox{$<$}} \break
                 \hbox{\lower 0.2\baselineskip \hbox{$\sim$}} } }

\begin{document}

\title{Heavy electrons from  Hund's rule and short-range Antiferromagnetism}
\author{Karyn Le Hur}
\affiliation{D\'epartement de Physique, Universit\'e de Sherbrooke, 
Sherbrooke, Qu\'ebec, Canada J1K 2R1}

\begin{abstract}
We investigate the one-dimensional ferromagnetic Kondo lattice in the hole-rich region. Of interest to us is the intermediate situation where the ferromagnetic Kondo coupling (Hund's coupling) is comparable to the electron bandwidth. The forced alignment favors triplet states whereas singlet states enter
in a quite low-density regime. The direct antiferromagnetic exchange between the core spins is assumed to still prevail over the double exchange. We discuss a solvable limit showing that short-range antiferromagnetism in the spin array will affect the coherent propagation of triplets, {\it i.e.}, turns a light triplet into a heavy singlet, resulting in a heavy electron ground state.

\end{abstract}
\pacs{71.10.-w,75.20.Hr,75.10.-b}

\maketitle

\section{Introduction}

The Kondo effect basically involves a localized magnetic impurity embedded in a bulk metal and antiferromagnetically coupled to conduction electrons. This strong-correlation phenomenon occurring in a variety of different systems and settings\cite{Hewson,Glazman,Manoharan,Zoler} really acts 
as a paradigm in the field of strongly-correlated systems. A fascinating ground state where the impurity is screened by a cloud of heavy electrons emerges at low temperatures.\cite{Nozieres} Even though the one-impurity Kondo effect is well understood\cite{Tsvelik} the practical situation of Kondo alloys with a finite concentration of impurities such as heavy fermion compounds is more subtle\cite{Emery} and especially when approaching a quantum critical point.\cite{Piers} At the heart of the central problem in those Kondo heavy fermion systems is the question how local moments behave itinerantly and in particular how the conduction electrons are counted in the presence of local spins. An attempt to study this question has been {\it e.g.} performed in Refs. \onlinecite{Oshikawa,Nolting}. Here, we
are rather concerned by the ferromagnetic region of the Kondo coupling. Of interest to us is the ferromagnetic Kondo lattice model in the hole-rich region. This model can be of relevance for a plethora of electron systems dominated by Hund's rule, such as oxide manganites\cite{Jin} or the heavy fermion metal LiV$_2$O$_4$.\cite{Kondo,Takagi}

Our starting point is the one-dimensional (1D) model,
\begin{eqnarray}
H &=&-t \sum_{\langle i,j\rangle,\alpha} c^{\dagger}_{i\alpha} c_{j\alpha} \\ \nonumber
&-& J_H\sum_{i,\alpha,\beta} c_{i\alpha}^{\dagger}\frac{\vec{\sigma}_{\alpha\beta}}{2}c_{i\beta}\vec{S}_i 
+J_{AFM}\sum_i \vec{S}_i\vec{S}_{i+1},
\end{eqnarray}
where $c_{i\alpha}^{\dagger}$ creates a conduction electron of spin $\alpha$ at the site $i$ and $\vec{S}_i$ depicts the spin $\frac{1}{2}$ local moments; the conduction band embodies a Fermi gas. 
The conduction electrons and the localized spins are coupled through a ferromagnetic Kondo coupling
or Hund's coupling and therefore $J_H>0$. The ``core'' spins are also directly coupled through an antiferromagnetic interaction $J_{AFM}$. 

 \begin{figure}[htbp]
\begin{center}
\includegraphics[width=8.7cm,height=3.3cm]{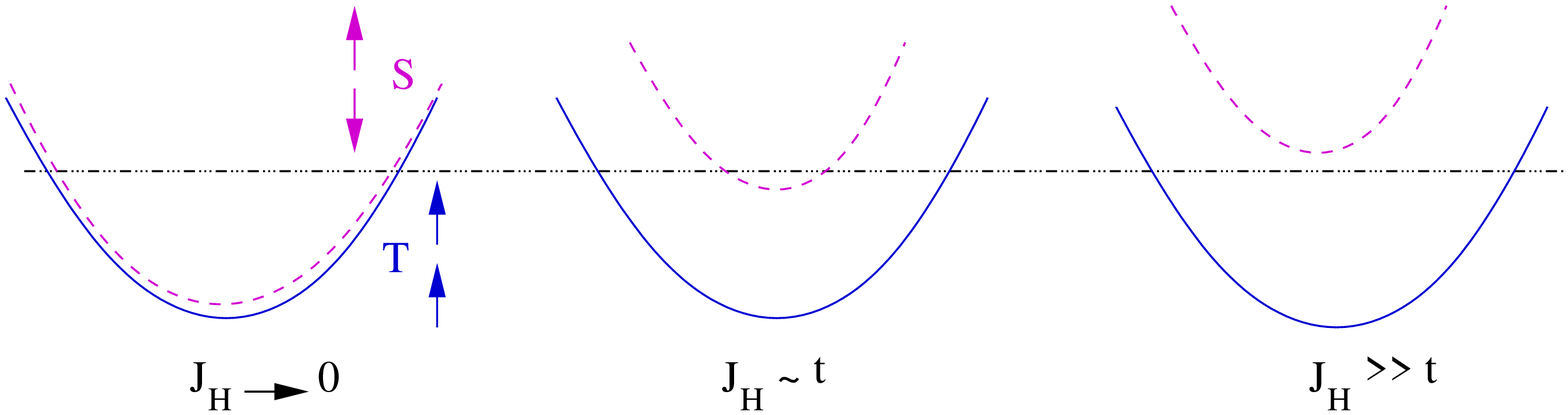}
\end{center}
\vskip -0.6cm
\caption{(color online) Schematic view of the depopulation effect of singlet states by increasing the ferromagnetic Kondo coupling. Here, we study the intermediate region $J_H\sim t$ where the singlet states enter in a low density limit  and weak antiferromagnetism between local moments still dominates.}
\label{setup}
\end{figure}

For $J_H\ll (J_{AFM},t)$, as a reminiscence of the single-impurity case, the ferromagnetic Kondo coupling fatally scales to very weak couplings at low temperatures and thus can be ignored; the conduction electrons essentially decouple from the spin array that forms a spin liquid with short-range magnetism due to reduced dimensionality.\cite{KLH}  

In contrast, in the large $J_H\gg t$ limit, the conduction electron spins adiabatically follow the core spins. As elucidated by Anderson and Hasegawa,\cite{Anderson} this induces 
a ferromagnetic correlation $J_{dex}[\frac{J_H}{t}]$ between neighbouring spins in order to facilitate coherent propagation. This phenomenon, known as the double exchange mechanism, leads to a ferromagnetic ground state at intermediate densities if we assume reasonable values of $J_{AFM}$ (for $J_H=8t$, following the Monte Carlo results of Ref. \onlinecite{Dagotto2} this precisely means $J_{AFM}<0.11 t$). On the other hand, the forced alignment also removes a large part of the Hilbert space since at each site, doubly-occupied and the antiparallel singly-occupied states are both projected out.  More precisely, an electron combines with the core spin to form two manifolds of total spin $S=0$ (singlet states) and $S=1$ (triplet states). In the large $J_H$ realm, singlet states are thus completely projected out as shown in Fig. 1. Moreover, in the ferromagnetic regime, the system can be identified as an ideal metal since triplets can propagate coherently, {\it i.e.}, still behave as free fermions.\cite{Betouras} 

Below, we are concerned by thermodynamic properties of conduction electrons in the intermediate region  $J_H\sim t$ where the double exchange becomes less important\cite{Dagotto,Betouras} producing $J_{AFM}>J_{dex}$ and thus a residual antiferromagnetic exchange $K=(J_{AFM}-J_{dex})>0$ but nevertheless singlet states enter in a low-density regime as a result of the strong Hund's coupling. In particular, one interesting question arises: does the presence of heavy singlets in low density and of the short-range antiferromagnetism in the spin array will hinder the coherent propagation of triplet states in the system and produce a {\it heavy electron ground state}? This is precisely the question addressed below that might be relevant to understand the heavy electron ground state in LiV$_2$O$_4$. Let us mention that the case where the singlet states are completely projected out  is beyond the scope of this paper and  we will not discuss the ferromagnetic phase.
Note that other scenarios involving an inter-site Kondo effect  (that might become important for larger Hund's couplings) have been explored in the context of LiV$_2$O$_4$.\cite{Rice,Coleman} 

\section{Model}

Again, the model under investigation is the Hamiltonian (1) in the intermediate realm where $J_H\sim t$
and $J_{AFM}>J_{dex}$ such that the core spins are coupled through a residual weak antiferromagnetic
interaction $K=(J_{AFM}-J_{dex})>0$ and form a low-dimensional spin liquid embodied by spin-1/2 (spinon) excitations.

We use the continuum limit and linearize the dispersion of the conduction electrons; thus, $c_{i\alpha}\rightarrow \sqrt{a}c_{\alpha}(x)$ (a being the lattice spacing) and $c_{\alpha}(x)$ is separated in left (-) movers $c_{-\alpha}(x)$ and right (+) movers $c_{+\alpha}(x)$. 

The spins $\vec{S}_i \rightarrow a\vec{S}(x)$ are related to the localized electrons via $\vec{S}(x)=\frac{1}{2}f^{\dagger}_{\alpha}(x)\vec{\sigma}_{\alpha\beta} f_{\beta}(x)$. We exploit the bosonization procedure for localized electrons $f_{p\alpha}(x)$:\cite{Karyn2} 
\begin{equation}
f_{p\alpha}(x)=[{\cal C}_p(x)z_{p\alpha}(x)]/\sqrt{2\pi a}. 
\end{equation}
Here, $p=\pm$ refers to the direction of propagation (right (+) or left (-)), $C_p(x)$ denotes the charge part and the spin (spinon) operator is precisely defined as in our Ref.\onlinecite{Karyn2}
\begin{equation}
z^{\dagger}_{p\alpha}(x)= \exp \left[i Q_s^{\alpha}\sqrt{\frac{\pi}{2}}(-p\phi_s+\theta_s)(x)\right].
\end{equation}
$Q_s^{\alpha}$ is related to the spin of a spinon via $S_z=Q_s^{\alpha}/2$ and $Q_s^{\alpha}=\pm 1$ for $\alpha=\uparrow,\downarrow$. Now, we can use the fact that charge fluctuations in the spin array are suppressed\cite{Karyn2} ${\cal C}^{\dagger}_+(x){\cal C}_{-}(x)=1$ as well as\cite{note0} $z^{\dagger}_{p\alpha}\vec{\sigma}_{\alpha\beta} z_{p\beta}{\cal C}^{\dagger}_p(x^+){\cal C}_{p}(x^-)=z^{\dagger}_{p\alpha}\vec{\sigma}_{\alpha\beta} z_{p\beta}[\frac{a}{a-ip(x^+-x^-)}]^{1/2}$ such that $\vec{S}(x)$ only depends on spinon operators. Note that deconfined spin 1/2 excitations have also been discussed in higher dimensions for certain classes of quantum critical magnets involving frustrated spin arrays\cite{Senthil} as well as for the Kondo lattice close to quantum criticality.\cite{pepin} In the context of LiV$_2$O$_4$ the lattice is definitely frustrated
ensuring the absence of any antiferromagnetic order. Moreover, short-range antiferromagnetism has been observed suggesting the formation of a spin-liquid state in the spin array.\cite{Kondo,Takagi}

 At this step, we apply the standard spin decomposition
\begin{equation}
\vec{S}(x)=\vec{{\cal L}}_-(x)+\vec{{\cal L}}_+(x) + (-1)^x\left(\vec{n}_-(x)+\vec{n}_+(x)\right),
\end{equation}
where $\vec{{\cal L}}_p(x) = \frac{1}{2\pi a}z^{\dagger}_{p\alpha}(x^+)\frac{\vec{\sigma}_{\alpha\beta}}{2}z_{p\beta}(x^-)
[\frac{a}{a-ip(x^+-x^-)}]^{1/2}$ 
embodies the $q=0$ {\it ferromagnetic} component whereas $\vec{n}_+= \frac{1}{2\pi a} z^{\dagger}_{+\alpha}\frac{\vec{\sigma}_{\alpha\beta}}{2} z_{-\beta}$ and $\vec{n}_-= \frac{1}{2\pi a} z^{\dagger}_{-\alpha}\frac{\vec{\sigma}_{\alpha\beta}}{2} z_{+\beta}$ stand for
the {\it staggered} magnetizations. Assuming incommensurate filling for the conduction band, hence we obtain
\begin{equation}
H=H_0 - {\cal J}_H \int dx \left(c^{\dagger}_{-\alpha}\frac{\vec{\sigma}_{\alpha\beta}}{2}
c_{-\beta}\right)\vec{{\cal L}}_- + (-\rightarrow +),
\end{equation}
where $H_0=H_{1DEG}+H_{spins}$ includes the kinetic energy of the conduction electrons as well the Heisenberg interaction between local moments yielding a usual plasmon form as a function of
$\phi_s$ and $\theta_s$,\cite{Karyn2} and ${\cal J}_H=J_H a>0$ depicts the dimensionless Hund's coupling. In particular, it is well established that the staggered magnetization of the local moments decouple from the conduction band at relatively (short) length scales $1/(a^{-1}\pi -2k_F)$ with $k_F$ being the Fermi momentum of the conduction band.\cite{KLH,Affleck}

Again, when $J_H\ll (K,t)$, under renormalization group (RG) flow, the ferromagnetic Kondo coupling would irrefutably flow to weak couplings\cite{KLH,Affleck} and the conduction band would essentially decouple from the spin array. Now, to judiciously tackle the non-perturbative region $J_H\sim t$ and $K\ll J_H$ where perturbative RG arguments cease to be valid, we resort to Eq. (2)  resulting in\cite{Shankar} ${\cal L}_-^+=[\exp(i\sqrt{2}\Phi_-)]/(2\pi a)$, ${\cal L}_+^+=[\exp(-i\sqrt{2}\Phi_+)]/(2\pi a)$, ${\cal L}_-^z=\partial_x\Phi_-/(2\sqrt{2}\pi)$, and ${\cal L}_+^z=\partial_x\Phi_+/(2\sqrt{2}\pi)$; we have introduced the chiral fields $\Phi_-=\sqrt{\pi}(\phi_s+\theta_s)$ and $\Phi_+=\sqrt{\pi}(\phi_s-\theta_s)$ and precise details of the calculations can be found in Appendix A. 

Our idea is to find a {\it solvable limit} analogous to the Toulouse limit for the single-impurity case\cite{Toulouse} that explicitly takes into account the singlet-triplet basis.

\section{Singlet-triplet mapping}

The Hund's coupling takes the form
\begin{eqnarray}
H_{{\cal J}_H} = \int dx \sum_{p=+,-} -\frac{{\cal J}_{H\perp}}{2\pi a} \left(e^{- ip\sqrt{2}\Phi_{p}}c^{\dagger}_{p\downarrow} c_{p\uparrow} +h.c.\right) \\ \nonumber
- \frac{{\cal J}_{Hz}}{4\sqrt{2}\pi} \nabla \Phi_{p} \left(c_{p\uparrow}^{\dagger} c_{p\uparrow}-c^{\dagger}_{p\downarrow}c_{p\downarrow}\right).
\end{eqnarray}
We can introduce composite fermionic objects by resorting to the precious ``gauge'' transformation
\begin{eqnarray}
{\cal K}_{\pm\uparrow} = c_{\pm\uparrow} \exp\left({\mp \frac{i}{\sqrt{2}}\Phi_{\pm}}\right) 
= c_{\pm\uparrow}z^{\dagger}_{\pm\uparrow}\\ \nonumber
{\cal K}_{\pm\downarrow} = c_{\pm\downarrow} \exp\left({\pm \frac{i}{\sqrt{2}}\Phi_{\pm}}\right)
= c_{\pm\downarrow}z^{\dagger}_{\pm\downarrow},
\end{eqnarray}
and the electron operators $c_{\pm\alpha}$ commute with the spinon operators. Exploiting the equality $c_{p\alpha}^{\dagger}c_{p\alpha}={\cal K}_{p\alpha}^{\dagger} {\cal K}_{p\alpha}$, we check
that the ${\cal K}$-objects are fermions: $\{{\cal K}_{p\uparrow}^{\dagger}(x),{\cal K}_{p\uparrow}(x)\}=1$. Additionally, ${\cal K}_{-\uparrow}(x){\cal K}_{-\uparrow}(y)=-c_{-\uparrow}(y)\exp(i\Phi_-(x)/\sqrt{2})c_{-\uparrow}(x)\exp(i\Phi_-(y)/\sqrt{2})$. After permuting $c_{-\uparrow}(x)\leftrightarrow c_{-\uparrow}(y)$, since the spin $\vec{S}(x)$ remains immobile, this explicitly produces the re-definitions ${\cal K}_{-\uparrow}(y)= c_{-\uparrow}(y)\exp(i\Phi_-(x)/\sqrt{2})$ and ${\cal K}_{-\uparrow}(x)= c_{-\uparrow}(x)\exp(i\Phi_-(y)/\sqrt{2})$. Hence, we can verify the fermionic  exchange statistics $\{{\cal K}_{-\uparrow}(x),{\cal K}_{-\uparrow}(y)\}=0$. 

The Hund's coupling then becomes
\begin{eqnarray}
H_{{\cal J}_H} = \int dx \sum_{p=+,-} -\frac{{\cal J}_{H\perp}}{2\pi a} \left({\cal K}^{\dagger}_{p\downarrow} {\cal K}_{p\uparrow} +h.c.\right) \\ \nonumber
- \frac{{\cal J}_{Hz}}{4\sqrt{2}\pi} \nabla \Phi_{p} \left({\cal K}_{p\uparrow}^{\dagger} {\cal K}_{p\uparrow}-{\cal K}^{\dagger}_{p\downarrow}{\cal K}_{p\downarrow}\right).
\end{eqnarray} 
We have decomposed the Hund's (Kondo) coupling ${\cal J}_H$ into an Ising $({\cal J}_{Hz})$ and a transverse part $({\cal J}_{H\perp})$.
Moreover, by linearizing the spectrum of the conduction electrons around the Fermi points, $H_{1DEG}$
turns into
\begin{eqnarray}
H_{1DEG} =  \int dx\ \sum_{\alpha=\uparrow,\downarrow} i v_F\left({\cal K}^{\dagger}_{-\alpha}\nabla {\cal K}_{-\alpha}-{\cal K}^{\dagger}_{+\alpha}\nabla {\cal K}_{+\alpha}\right) &&\\ \nonumber
+\sum_{p=\pm}\frac{v_F}{\sqrt{2}}\nabla \Phi_{p} \left({\cal K}_{p\uparrow}^{\dagger} {\cal K}_{p\uparrow}-{\cal K}^{\dagger}_{p\downarrow}{\cal K}_{p\downarrow}\right), &&
\end{eqnarray}
where $v_F=2ta\sin(k_Fa)$ is the Fermi velocity. {\it We identify a solvable point ${\cal J}_{Hz}=4\pi v_F$}. Indeed, for ${\cal J}_{Hz}=4\pi v_F$, we can easily reveal the singlet-triplet basis by resorting to the symmetric and antisymmetric combinations, ${\cal T}_{p}=({\cal K}_{p\uparrow}+{\cal K}_{p\downarrow})/\sqrt{2}$ (triplets) and  ${\cal S}_{p}=({\cal K}_{p\uparrow}-{\cal K}_{p\downarrow})/\sqrt{2}$ (singlets). More precisely, we get
\begin{eqnarray}
H= \int dx\ i v_F\left({\cal T}_-^{\dagger}\nabla{\cal T}_-  -{\cal T}^{\dagger}_+\nabla {\cal T}_+\right)
+({\cal T}\rightarrow {\cal S})\\ \nonumber
-\sum_{p=+,-} \frac{{\cal J}_{H\perp}}{2\pi a} \left({\cal T}^{\dagger}_{p} {\cal T}_{p} -{\cal S}_{p}^{\dagger}{\cal S}_{p}\right) + H_{spins}.
 \end{eqnarray}
The Hamiltonian (10) suggests the following interpretation: ${\cal T}$ refers to an itinerant electron with the lowest energy, {\it i.e.}, forming a {\it triplet} state with the local moment at the same site (that is in agreement with the strong ${\cal J}_{Hz}$ and ${\cal J}_{H\perp}$ limit) whereas ${\cal S}$ refers to an electron forming a {\it singlet} state with the local moment at the same site. A singlet costs a supplementary energy $\Delta={\cal J}_{H\perp}/a\pi$ as it should be because this requires the spin flip of a conduction electron or of a local moment. Note, the weak magnetism of the spin array is preserved through $H_{spins}$ as it should be for ferromagnetic Kondo couplings.  Note also that the ${\cal K}_{p\uparrow}$ operators do not have a simple interpretation due to the coupling term $-\frac{{\cal J}_{H\perp}}{2\pi a} \left({\cal K}^{\dagger}_{p\downarrow} {\cal K}_{p\uparrow} +h.c.\right)$; however, it is essential to apply the gauge transformation of Eq. (7) to reveal the (physical) singlet-triplet basis.

\section{Low-density singlet regime}

Eq. (10) ensures the following Fermi wave vectors:
 \begin{equation}
 -2t\cos(k_F^{{\cal T},{\cal S}}a) \mp \frac{{\cal J}_{H\perp}}{2\pi a} = -2t\cos(k_F a).
 \end{equation}
 Having in mind LiV$_2$O$_4$, we envision the situation of a quarter-filled band. 
We infer the renormalized (Fermi) velocities $v_{{\cal T},{\cal S}}\simeq 2ta(1-\left[\cos(k_F a)\mp \Delta/(4 t)\right]^2)^{1/2}$.  For $\Delta/t\rightarrow 0$, we can check that  the Hund's coupling has practically no effect, $v_{\cal S}=v_{\cal T}\approx v_F$.  In contrast, for extremely large Hund's couplings such that $\Delta>2t[2-\sqrt{2}]$ the singlet band becomes completely depopulated and $v_{\cal S}=0$. 
We consider the intermediate coupling region where singlets occur in low density: $0<v_{\cal S}\ll v_{\cal T}\sim v_F$.

\subsection{Solvable limit: ${\cal J}_{Hz}=4\pi v_F$}

Let us first discuss thermodynamic properties of the whole system at the solvable limit ${\cal J}_{Hz}=4\pi v_F$ where itinerant electrons and spinons {\it essentially} decouple; see Eq. (10). Temperature is assumed to be smaller than $K$.

First, through $H_{spins}$ the Curie-Weiss susceptibility of the spin array will saturate to $\chi_{K}=L/(2\pi v_K)$ and the specific heat of the spin array reads $C_K=\pi TL/(3v_K)$.\cite{Giamarchi} Here,
$v_K=K/a$ embodies the velocity of the spinons.

Second, from Eq. (10), at ${\cal J}_{Hz}=4\pi v_F$, 
the electronic contribution to the specific heat takes the form $C_{el}=TL\pi[1/(3v_{\cal T})+1/(3v_{\cal S})]$. Below, we will thus refer to 
\begin{equation}
C_o =C_{el}+C_K \approx \frac{\pi T L}{3 v_K} + \frac{\pi LT}{3 v_{\cal S}},
\end{equation}
as the {\it total} specific heat at ${\cal J}_{Hz}=4\pi v_F$ with $0<v_{\cal S}\ll v_{\cal T}$. We can evaluate
the electronic part of the magnetic susceptibility by adding a term in the Hamiltonian (10):
\begin{eqnarray}
H_{\cal H} &=& -\frac{{\cal H}}{2}\sum_{p=\pm} \int dx \left(c^{\dagger}_{p\uparrow}c_{p\uparrow}-
c^{\dagger}_{p\downarrow}c_{p\downarrow}\right), \\ \nonumber
&=& -\frac{{\cal H}}{2}\sum_{p=\pm} \int dx \left({\cal S}^{\dagger}_{p}{\cal T}_{p}+{\cal T}^{\dagger}_{p}{\cal S}_{p}\right).
\end{eqnarray}
To align itinerant electron spins along the magnetic field direction
this inevitably requires some spin-flip processes, {\it i.e.}, some triplet-singlet transitions and vice-versa. To compute  the free energy contribution to second order in ${\cal H}$, $F_{\cal H}=-\frac{1}{2}\int_0^{\beta} d\tau\  \langle T H_{\cal H}(\tau) H_{\cal H}(0) \rangle$, we will exploit the Green function of the ${\cal T}_{\pm}$ itinerant particles 
\begin{equation}
G_{{\cal T}_{\pm}}(\tau,x-x') = \frac{1}{2\pi v_{\cal T}} \frac{\pi/\beta}{\sin\frac{\pi}{\beta}[\tau\mp i(x-x')/v_{\cal T}]},
\end{equation}
and similarly for ${\cal S}_{\pm}$. Hence, we obtain the expression
\begin{eqnarray}
F_{\cal H} = -\frac{{\cal H}^2 L}{4}\int_0^{\beta} d\tau \int d\bar{x} \ \left(\frac{1}{2\pi v_{\cal T}}\right)\left(\frac{1}{2\pi v_{\cal S}}\right)\left(\frac{\pi}{\beta}\right)^2\\ \nonumber
\times \left(\sum_{p=\pm} \frac{1}{\sin\left[\frac{\pi}{\beta}\left(\tau-ip\frac{\bar{x}}{v_{\cal T}}\right)\right]\sin\left[\frac{\pi}{\beta}\left(\tau-ip\frac{\bar{x}}{v_{\cal S}}\right)\right]}\right).
\end{eqnarray}
$\bar{x}=x-x'$ is the relative coordinate. The main contribution stems from $\bar{x}\approx 0$ and small $\tau$. 
 \begin{figure}[htbp]
\begin{center}
\includegraphics[width=8.6cm,height=2.9cm]{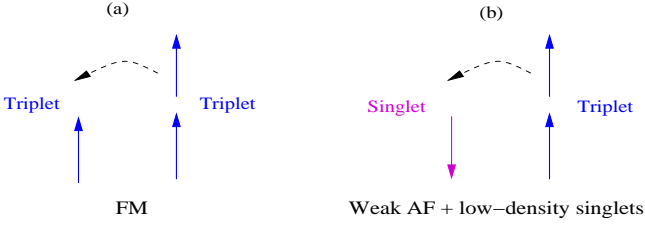}
\end{center}
\vskip -0.6cm
\caption{(color online) When the spin array is ferromagnetically ordered, triplets can freely propagate
 through the structure (a). When short-range antiferromagnetic correlations between the
core spins dominate, the hopping of conduction electrons from site to site will turn light triplets into heavy singlets (b) thus resulting in a heavy electron ground state.}
\label{setup}
\end{figure}
We introduce  the short-time cutoff $\tau_{min} = (2\pi \Delta)^{-1} \sim a/(2\pi v_F)$
such that
\begin{equation}
F_{\cal H} \approx -\frac{{\cal H}^2}{2}\frac{La \tau_{min}^{-1}}{4\pi^2 v_{\cal T} v_{\cal S}} = -\frac{L v_F {\cal H}^2}{4\pi v_{\cal S} v_{\cal T}}.
\end{equation}
This gives rise to the magnetic susceptibility $\chi_{el}=-\partial^2 F_{\cal H}/\partial {\cal H}^2 \approx Lv_F/(2\pi {v_{\cal S}}v_{\cal T})$. Note in passing that when $\Delta/t\rightarrow 0$, we recover
the free electron gas behavior, $C_{el}=2\pi TL/(3 v_F)$ and $\chi_{el}=L/(2\pi v_F)$. In the region of interest to us $0<v_{\cal S}\ll v_{\cal T}\sim v_F$, we rather obtain
\begin{equation}
\chi_o = \chi_{el}+\chi_K \approx \frac{L}{2\pi v_K} + \frac{L}{2\pi  {v_{\cal S}}}.
\end{equation}

The singlet velocity $v_{\cal S}$ mainly controls the thermodynamic properties of the conduction band for ${\cal J}_{Hz}=4\pi v_F$.

\subsection{Deviation from the solvable limit}

{\it We argue that there is a more important contribution induced by any realistic deviation from ${\cal J}_{Hz}=4\pi v_F$ that will lead to the emergent heavy fermion behavior. }

We put $\eta=\left(v_F-{\cal J}_{Hz}/4\pi\right)/\sqrt{2}>0$, resulting in
\begin{eqnarray}
H_{\eta} = \eta \int dx\ \sum_{p=\pm} \nabla\Phi_{p}\left({\cal S}^{\dagger}_{p}{\cal T}_{p}+h.c.\right).
\end{eqnarray}
We emphasize that this term has a clear physical meaning as illustrated in Fig. 2. More precisely, the short-range antiferromagnetism in the spin array will hinder the coherent propagation of triplets, {\it i.e.},  converts triplets into heavy singlets through the hopping of conduction electrons from site to site. Therefore, this should strongly affect thermodynamic properties of the system. Now, to show this explicitly we shall compute the free energy to second order in $\eta$
and in particular exploit the spinon Green function $(G_{\nabla\Phi_p}(\tau,\bar{x})=\langle \nabla\Phi_{p}(\tau,\bar{x})\nabla\Phi_{p}(0,0)\rangle)$\cite{Giamarchi}
\begin{equation}
G_{\nabla\Phi_{\pm}}(\tau,\bar{x}) =2\pi^2 \left(\frac{(2v_K\beta)^{-1}}{\sin\frac{\pi}{\beta}[\tau \mp i\bar{x}/v_{K}]}\right)^2.
\end{equation}
We extract two distinct contributions to the free energy
\begin{eqnarray}
F_1 &=& -2\eta^2 (La) \int_{0}^{\beta} d\tau \left[G_{{\cal T}+} G_{{\cal S}+}G_{\nabla\Phi_+}\right](\tau,0)\\ \nonumber
F_2 &=& -\frac{\eta^2}{2} (La) \sum_{p=\pm} \langle {\cal S}^{\dagger}_{p}{\cal T}_{p}+h.c.\rangle^2 \int_0^{\beta} d\tau\ G_{\nabla\Phi_+}(\tau,0).
\end{eqnarray}
Since those terms involve the core spins the short-time cutoff has to be modified
accordingly: $\tau_{min}\sim 1/(2\pi K)$. Moreover, by exploiting Eq. (13), we  identify $\langle {\cal S}^{\dagger}_{p}{\cal T}_{p}+h.c.\rangle=\chi_{el} {\cal H}/L \approx {\cal H}/(2\pi v_{\cal S})$. Finally, we obtain:\cite{note}
\begin{eqnarray}
F_1 &=& - 2\eta^2 T^2 \frac{L\pi}{6 v_{\cal S} v_{\cal T} v_K} \\ \nonumber
F_2 &=& - 2\eta^2 {\cal H}^2 \frac{L}{2\pi v_{\cal S}^2 4v_K}.
\end{eqnarray} 
We deduce that $F_1$ will (deeply) modify the specific heat whereas $F_2$ will renormalize the magnetic susceptibility:
\begin{eqnarray}
\delta C &=& C_{o} \frac{{v_F}}{v_{\cal S}+v_K}{\left[1-{\cal J}_{Hz}/(4\pi v_F)\right]^2}\\ \nonumber
\delta \chi &=& \chi_{o} \frac{{v_F}^2}{2 v_{\cal S}(v_K+v_{\cal S})}\left[1-{\cal J}_{Hz}/(4\pi v_F)\right]^2.
\end{eqnarray}
Since $(v_K,v_{\cal S})\ll v_F$, we find that $\delta C\gg C_o$ and
$\delta\chi \gg \chi_o$. {\it It is relevant to note that the emergent heavy fermion scale is $\sim\eta/a$}. We infer the following Wilson ratio
\begin{equation}
R_w=\frac{\delta\chi/\chi_o}{\delta C/C_o}\approx\frac{v_F}{2v_{\cal S}}.
\end{equation}
Experiments on LiV$_2$O$_4$ \cite{Kondo,Takagi}  report $R_w=1.8$ suggesting $v_F\approx 3.6 v_{\cal S}$. Remember that the heavy fermion metal emerges due to the prolific combination between the spin environment with short-range antiferromagnetism and the prominent Hund's coupling  leading to  $(v_K,v_{\cal S})\ll v_F$. 

\section{Conclusion}

To summarize, we have studied thermodynamic properties of conduction electrons in the intermediate region $J_H\sim t$ of the ferromagnetic Kondo lattice model where short-range antiferromagnetic correlations between the core spins have been assumed to still prevail over the ferromagnetic double exchange $J_{dex}$. Singlet states enter in a quite low density regime as a result of the prominent Hund's coupling. We have explored a solvable limit showing explicitly that  short-range antiferromagnetic correlations between the core spins affect the coherent propagation of triplet states, {\it i.e.}, convert a light triplet into a heavy singlet, resulting in a heavy fermion ground state. The situation is distinguishable from the strong Hund's coupling phase with ferromagnetic ordering where triplets can propagate coherently and thus behave as free fermions.\cite{Betouras} This work might be relevant to explain the heavy fermion physics of LiV$_2$O$_4$\cite{Kondo,Takagi} without invoking Kondo physics between a core spin and a conduction electron on a neighboring site.\cite{Rice,Coleman}  It would be useful to develop similar theories in higher dimensions.

{\it Acknowledgements.---} K.L.H. thanks M. Rice for discussions and KITP (Santa-Barbara) via the ``Quantum Phase Transition workshop'' (2005, NSF PHY99-07949). This work was supported by FQRNT, CIAR, NSERC.

\begin{appendix}

\section{Bosonization dictionary}

Here, we provide a pedestrian derivation of the $q=0$ component of  the spin operator $\vec{S}(x)$ starting from
\begin{equation}
f_{p\alpha}^{\dagger}(x)=[{\cal C}_p^{\dagger}(x)z_{p\alpha}^{\dagger}(x)]/\sqrt{2\pi a}.
\end{equation}
We have introduced the spin (spinon) operator of Eq. (3) and the charge operator. Since charge fluctuations are completely suppressed in the spin array one obtains 
\begin{equation}
\vec{{\cal L}}_p(x) = \frac{1}{2\pi a}z^{\dagger}_{p\alpha}(x^+)\frac{\vec{\sigma}_{\alpha\beta}}{2}z_{p\beta}(x^-)
\sqrt{\frac{a}{a-ip(x^+-x^-)}},
\end{equation}
as explained in page 1. We can evaluate ${\cal L}_+^+$ exploiting
\begin{equation}
z^{\dagger}_{+\uparrow}(x^+)z_{+\downarrow}(x^-)= e^{i\sqrt{2\pi}(-\phi_s+\theta_s)}\sqrt{\frac{a}{a-i(x^+-x^-)}}.
\end{equation}
This results in
\begin{eqnarray}
{\cal L}_+^+ &=& \frac{1}{2\pi a} e^{i\sqrt{2\pi}(-\phi_s+\theta_s)} \frac{a}{a-i(x^+-x^-)} \\ \nonumber
&\rightarrow& \frac{1}{2\pi a} e^{-i\sqrt{2}\Phi_+},
\end{eqnarray}
and we have defined $\Phi_+=\sqrt{\pi}(\phi_s-\theta_s)$. We also infer
\begin{eqnarray}
f^{\dagger}_{+\uparrow} f_{+\uparrow} &=& -\frac{i}{2\sqrt{2}\pi a}\partial_x\Phi_+ \frac{a(x^+-x^-)}{a-i(x^+-x^-)}\\ \nonumber
&\sim&  \frac{1}{2\sqrt{2}\pi}\partial_x\Phi_+.
\end{eqnarray}
In a similar way we get
\begin{eqnarray}
f^{\dagger}_{+\downarrow} f_{+\downarrow} &=& \frac{i}{2\sqrt{2}\pi a}\partial_x\Phi_+ \frac{a(x^+-x^-)}{a-i(x^+-x^-)}\\ \nonumber
&\sim&  -\frac{1}{2\sqrt{2}\pi}\partial_x\Phi_+,
\end{eqnarray}
and therefore ${\cal L}_+^z = \frac{1}{2\sqrt{2}\pi}\partial_x\Phi_+$. We also extract
\begin{eqnarray}
{\cal L}_-^+ &=& \frac{1}{2\pi a} e^{i\sqrt{2\pi}(\phi_s+\theta_s)} \frac{a}{a+i(x^+-x^-)} \\ \nonumber
&\rightarrow& \frac{1}{2\pi a} e^{i\sqrt{2}\Phi_-},
\end{eqnarray}
where $\Phi_-=\sqrt{\pi}(\phi_s+\theta_s)$. Finally, we find
\begin{eqnarray}
{\cal L}_-^z &=& \frac{1}{2}(f^{\dagger}_{-\uparrow} f_{-\uparrow} - f^{\dagger}_{-\downarrow} f_{-\downarrow})\\ \nonumber
&=& \frac{1}{2\sqrt{2}\pi}\partial_x\Phi_-.
\end{eqnarray}
This is in accordance with the definitions of Ref. \onlinecite{Giamarchi}.

\end{appendix}


\begin{thebibliography}{10}
\bibitem{note0}
Even though $[a/(a-ip(x^+-x^-))]^{1/2}$ tends to 1 $(x_+\sim x_-)$ 
we keep it explicitly to obtain the correct ${\cal L}_p^z$ in page 1.\cite{Shankar}
\bibitem{note}
$F_1\propto T^2$ and $F_2\propto {\cal H}^2$ are in agreement with the fact that
\vskip -0.05cm
$\eta$ is a marginal coupling.
\end{thebibliography}
\end{document}